\documentstyle[12pt]{article}
\def\PRL #1 #2 #3 {Phys.~Rev.~Lett.~{\bf #1}, #2 (#3)}
\def\PRD #1 #2 #3 {Phys.~Rev.~D~{\bf #1}, #2 (#3)}
\def\PLB #1 #2 #3 {Phys.~Lett.~B~{\bf #1}, #2 (#3)}
\def\NPB #1 #2 #3 {Nucl.~Phys.~{\bf B#1}, #2 (#3)}

\newcommand{\mysectionstar}[1]{
\par\bigskip\noindent{\large\bf #1}\nopagebreak[4]\par\vskip .3cm}

\begin{document}
\include{psfig}
\begin{titlepage}

\rightline{hep-ph/9710492}
\medskip
\rightline{October 1997}
\bigskip\bigskip
\begin{center} {\Large \bf Group-Theoretic Evidence for \\
\medskip
SO(10) Grand Unification} \\
\medskip
\bigskip\bigskip\bigskip\bigskip
{\large{\bf J.~Lykken}} \\ 
\medskip 
Fermi National Accelerator Laboratory \\
P.~O.~Box 500 \\ Batavia, IL\ \ 60510 \\  
\bigskip\bigskip
{\large{\bf T.~Montroy}\footnote{Present address: Department of Physics,
University of California, Santa Barbara, CA 93106}
and {\bf S.~Willenbrock}} \\ 
\medskip 
Department of Physics \\
University of Illinois \\ 1110 West Green Street \\  Urbana, IL\ \ 61801 \\
\bigskip 
\end{center} 
\bigskip\bigskip\bigskip

\begin{abstract}
The hypercharges of the fermions are not uniquely determined in SO(10) grand
unification, but rather depend upon which linear combination of the two U(1)
subgroups of SO(10)$\supset\,$SU(3)$\times$SU(2)$\times$U(1)$\times$U(1)
remains unbroken. We show that, in general, a given hypercharge assignment
can be obtained only with very high-dimensional Higgs representations.
The observation that the standard model is obtained with low-dimensional
Higgs representations can therefore be regarded as further evidence for
SO(10) grand unification. This evidence is independent of the fact that
SO(10)$\supset\,$SU(5).
\end{abstract}

\end{titlepage}

\newpage

\indent\indent The standard model of the fundamental interactions is based 
on the gauge group 
SU(3)$_c\times$SU(2)$_L\times$U(1)$_Y$.  Each generation of fermions 
transforms as a reducible representation of the gauge group, consisting 
of five irreducible representations.\footnote{This does not include a 
right-handed neutrino.  We shall introduce this particle into the 
discussion shortly.}  
The hypercharges of the representations are chosen to reproduce 
the observed electric charges of the fermions.

One of the most compelling pieces of evidence for grand unification is 
that the fermions of each generation transform as the $\overline 5$ + 
10-dimensional representation of SU(5) \cite{GG}.
The five irreducible representations of each generation of fermions are 
thereby unified into two, and the three gauge groups are unified into one.  
The hypercharges of the five irreducible representations 
are uniquely determined by their embedding in the $\overline 5$ + 10 of SU(5).

If a right-handed neutrino exists, the group-theoretic 
evidence for grand unification is
even more compelling: the fermions of each generation transform as the 
16-dimensional representation of SO(10) \cite{G,FM}.  The six irreducible
representations are thereby unified into a single irreducible representation,
and the three gauge groups are unified into one.  We assume the existence 
of a right-handed neutrino for the remainder of the discussion.

SO(10) has a subgroup SU(3)$\times$SU(2)$\times$U(1)$\times$U(1).
When SO(10) is spontaneously broken to SU(3)$_c\times$SU(2)$_L\times$U(1)$_Y$,
the hypercharge subgroup\footnote{The hypercharge subgroup is, by definition,
the unbroken U(1) subgroup.  In general, this does not correspond to the 
usual hypercharge subgroup of the standard model.} 
is a linear combination of the two U(1) subgroups
of SO(10).  Thus the hypercharges of the 
fermions are not uniquely determined in SO(10) grand unification, in contrast
to the case of SU(5) grand unification, but rather
depend upon which linear combination of the two U(1) subgroups is unbroken.

The SU(3)$_c\times$SU(2)$_L\times$U(1)$_Y$
quantum numbers of the left-handed fields which make up the 16-dimensional 
representation of SO(10) are given in Table 1.
The hypercharge is normalized such that the left-handed 
positron has unit hypercharge.  The parameter $a$ 
depends upon which linear combination of the two U(1) subgroups is 
unbroken. It is a rational number because the hypercharges are ``quantized'', 
i.e., commensurate, since a U(1) subgroup of a non-Abelian group is 
necessarily compact \cite{GG}.\footnote{This representation is free of 
gauge and gravitational anomalies for any value of $a$, not just
for rational values \cite{GM,BM,FJLV,MRW}.   
It thus serves as an example of a chiral, anomaly-free
gauge theory that (for irrational values of $a$) cannot be embedded in a 
grand-unified theory \cite{G2}.} 

\begin{table}[htb]
\begin{center}
\caption[fake]{Quantum numbers of the left-handed fields which make up the 
16-dimensional representation of SO(10).  The parameter $a$ depends upon 
how the U(1)$_Y$ subgroup is embedded in SO(10).}
\bigskip
\begin{tabular}{ccccc}
  & SU(3)$_c$ & SU(2)$_L$ & U(1)$_Y$ & U(1)$_{EM}$\\
\\
$(u_L,d_L)$     & 3 & 2 & $a$ & $(1-2a, 4a-1)$ \vspace{4pt}\\
$u_L^c$   & $\overline 3$ & 1 & $2a-1$ & $2a-1$ \vspace{4pt}\\
$d_L^c$   & $\overline 3$ & 1 & $1-4a$ & $1-4a$ \vspace{4pt}\\
$(\nu_L,e_L)$ & 1 & 2 & $-3a$ & $(1-6a, -1)$ \vspace{4pt}\\
$\nu_L^c$ & 1 & 1 & $6a-1$ & $6a-1$ \vspace{4pt}\\
$e_L^c$   & 1 & 1 & 1 & 1 \\
\end{tabular}
\end{center}
\end{table}

The value of the parameter $a$ depends upon the Higgs 
representation employed to break SO(10) to 
SU(3)$_c\times$SU(2)$_L\times$U(1)$_Y$. The Higgs field may be
either fundamental or composite; only its group-theoretic properties are
relevant to the considerations of this paper.
The candidate values of $a$ for a given irreducible representation 
correspond to the SU(3)$\times$SU(2)$\times$U(1) singlets contained 
in that representation \cite{S}.  Usually this 
representation must be accompanied by at least one additional 
Higgs irreducible representation 
in order to break SO(10) down to SU(3)$\times$SU(2)$\times$U(1), because the 
latter is generally not a maximal little group of the former for a single 
irreducible representation \cite{S}.
To generate fermion masses, the SU(2)$_L\times$U(1)$_Y$ symmetry must be 
broken by yet one or more additional Higgs irreducible representation, 
chosen from the 10-, 120-, and 126-dimensional representations (since 
$16\times 16 = 10 + 120 + 126$).  The SU(2)$_L\times$U(1)$_Y$ symmetry
is broken to U(1)$_{EM}$ when any of the color-singlet, SU(2) doublets 
contained in these
representations acquires a vacuum-expectation value, leading to the 
electric charges listed in the last column of Table 1.
The standard model evidently corresponds to $a=1/6$.

We have shown by construction that any rational value of $a$ can be obtained
by an appropriate choice of the Higgs irreducible representation.  However,
a given value of $a$ generally requires a very large Higgs irreducible 
representation.
In practice, the smallest Higgs irreducible representations yield only a 
few values of $a$.
We list in Table 2 the possible values of $a$, and the Higgs 
irreducible representations which can yield that value, for all Higgs 
representations 
of dimension 55440 or less.\footnote{These were derived with the help of 
the tables of Refs.~\cite{S,MP} and the
{\it Liegroup} package developed by George M. Hockney.}
Note that $a$ and $a/(6a-1)$ are equivalent, upon 
interchanging $u_L^c$ with $d_L^c$ and $\nu_L^c$ with $e_L^c$.  Thus we list
values of $a$ in the interval $[0,1/3]$ only. 
Higgs representations listed 
as ``undetermined'' have SU(3)$\times$SU(2)$\times$U(1)$\times$U(1) singlets, 
which do not determine $a$.  Higgs representations listed as ``none''
have no SU(3)$\times$SU(2)$\times$U(1) singlets.  If one or more additional
Higgs irreducible representations are needed to break SO(10) to 
SU(3)$\times$SU(2)$\times$U(1), as is generally the case,
they must correspond to the same value of $a$ or an undetermined
value of $a$.

It is satisfying that the standard model 
($a=1/6$) is obtained with several small Higgs irreducible
representations,\footnote{Generally accompanied by at 
least one other irreducible representation, such as the 
45-dimensional representation \cite{R}.}
the 16-, 126-, and 144-dimensional representations, as is well known \cite{R}.
If we lived in a world in which the ratio of
the hypercharges of the quark doublet and the positron were, say, 
1/8 rather than
1/6, we could still embed the fermions in the 16-dimensional representation
of SO(10), but we would need a 9504-dimensional Higgs representation to
obtain the desired symmetry breaking.  While there is 
(perhaps) nothing fundamentally
wrong with this, it is less palatable than a model which requires only Higgs
fields in low-dimensional irreducible representations.
These results are independent of the fact that 
SO(10)$\supset\,$SU(5); for example the 144-dimensional representation, 
which contains no SU(5)
singlet, produces $a=1/6$, while the 210-dimensional representation, 
which does contain an SU(5) singlet, produces $a=1/3$.
  
We believe that the economy
of the Higgs representation in SO(10) grand unification, while well known,
has not been fully appreciated. We regard it as further evidence
for SO(10) grand unification.

\mysectionstar{Acknowledgements}

\indent \indent We are grateful for conversations with G.~Anderson, D.~Berg, 
R.~Leigh, A.~Nelson, and P.~Ramond.  S.~W.~thanks the Aspen Center for 
Physics, where part of this work was performed.
The research of J.~L.~was supported by the Fermi National
Accelerator Laboratory, which is operated by Universities Research
Association, Inc., under contract no. DE-AC02-76CHO3000.
S.~W.~and T.~M.~were supported in part by Department of Energy grant 
DE-FG02-91ER40677.  T.~M.~was supported in part by the Lorella M.~Jones 
UIUC Summer Research Fellowship in Physics.

\newpage
\thispagestyle{empty}

\begin{table}[htb]
\begin{center}
\caption[fake]{Values of the parameter $a$ corresponding to the 
SU(3)$\times$SU(2)$\times$U(1) singlets of SO(10) Higgs representations 
up to dimension 55440.  The standard model corresponds to 
$a=1/6$.  The representations labeled ``undetermined''
have SU(3)$\times$SU(2)$\times$U(1)$\times$U(1) singlets, and those labeled
``none'' have no SU(3)$\times$SU(2)$\times$U(1) singlets.}
\bigskip

\begin{tabular}{cl}
$a$ & {\rm SO(10) Higgs representation}             \\
\\
1/6 & 16, 126, 144, 560, 672, 720, 1200, 1440, 1728, 2640, 2772, 2970, 3696, \\
    & 3696$^\prime$, 4950, 5280, 6930$^\prime$, 7920, 8064, 8800, 9504, 10560, 
      11088, 15120, \\ 
    & 17280, 20592, 20790, 23760, 25200, 26400, 27720, 28160,
      28314, 29568, \\ 
    & 30800, 34398, 34992, 36750, 38016, 39600, 43680, 46800, 
      48048, 48048$^\prime$, \\
    & 48114, 49280, 50050, 50688, 55440 \\ 
\\
1/3 & 45, 210, 770, 945, 1050, 1386, 4125, 5940, 6930, 7644, 8085, 8910, 
      12870, \\
    & 14784, 16380, 17325, 17920, 23040, 50688, 52920 \\ 
\\
0   & 120, 126, 1728, 2772, 2970, 3696$^\prime$, 4125, 4312, 4950, 6930, 
      6930$^\prime$, 10560, \\
    & 20790, 27720, 28160, 28314, 34398, 36750, 42120, 46800, 48114, 50050, \\
    & 50688 \\ 
\\
1/4 & 560, 1440, 3696, 5280, 8064, 8800, 11088, 15120, 23760,
      25200, 29568, \\
    & 30800, 34992, 38016, 39600, 43680, 46800, 48048, 49280,
      55440 \\ 
\\
1/12& 672, 1200, 8800, 11088, 17280, 23760, 25200, 26400, 28314, 30800, \\ 
    & 34992, 38016, 49280, 55440 \\ 
\\
1/9 & 2772, 6930, 50688 \\ 
\\
2/9 & 3696$^\prime$, 6930$^\prime$, 20790, 34398, 36750, 46800, 48114 \\ 
\\
5/18& 8064, 34992, 39600, 43680 \\ 
\\
1/8 & 9504, 29568 \\ 
\\
1/18& 9504, 29568, 30800 \\ 
\\
5/24& 17280, 26400 \\ 
\\
2/15& 28314 \\
\\
undetermined & 45, 54, 210, 660, 770, 945, 1050, 1386, 4125, 4290, 5940, 6930,
               7644, 8085,\\
    &          8910, 12870, 14784, 16380, 17325, 17920, 19305, 23040, 
               50688, 52920\\
\\
none& 10, 210$^\prime$, 320, 1782, 4410, 4608, 9438, 31680, 37180, 37632,
      48510\\ 
\\
\end{tabular}
\end{center}
\end{table}

\end{document}